\documentclass{article}

\usepackage{gensymb}
\usepackage{PRIMEarxiv}

\usepackage[utf8]{inputenc} % allow utf-8 input
\usepackage[T1]{fontenc}    % use 8-bit T1 fonts
\usepackage{hyperref}       % hyperlinks
\usepackage{url}            % simple URL typesetting
\usepackage{booktabs}       % professional-quality tables
\usepackage{amsfonts}       % blackboard math symbols
\usepackage{nicefrac}       % compact symbols for 1/2, etc.
\usepackage{microtype}      % microtypography
\usepackage{lipsum}
\usepackage{fancyhdr}       % header
\usepackage{graphicx}       % graphics
\graphicspath{{media/}}     % organize your images and other figures under media/ folder

\title{DFT analysis and demonstration of enhanced clamped Electro-Optic tensor by strain engineering in PZT}

\author{
  Suraj, Shankar Kumar Selvaraja \\
  Center for Nanoscience and Engineering \\
  Indian Institute of Science, Bangalore 560012, India \\
  \texttt{\{Suraj,Shankar Kumar Selvaraja\} suraj2@iisc.ac.in, shankarks@iisc.ac.in} \\
}

\begin{document}
\maketitle

\vspace{-2em}
\begin{abstract}
We report $\approx$400\% enhancement in PZT Pockels coefficient on DFT simulation of lattice strain due to phonon mode softening.The simulation showed a relation between the rumpling and the Pockels coefficient divergence that happens at -8\% and 25\% strain developed in PZT film. The simulation was verified experimentally by RF sputter deposited PZT film on Pt/SiO$_2$/Si layer. The strain developed in PZT varied from -0.04\% for film annealed at 530\degree C to -0.21\% for 600\degree C annealing temperature. The strain was insensitive to RF power with a value of -0.13\% for power varying between 70-130 W. Pockels coefficient enhancement was experimentally confirmed by Si Mach–Zehnder interferometer (MZI) loaded with PZT and probed with the co-planar electrode. An enhancement of $\approx$300\% in Pockels coefficient was observed from 2-8 pm/V with strain increasing from -0.04\% to -0.21\%. To the best of our knowledge, this is the first time study and demonstration of strain engineering on Pockels coefficient of PZT using DFT simulation, film deposition, and photonic device fabrication.  

\end{abstract}

% \setboolean{displaycopyright}{false} % Do not include copyright or licensing information in submission.
\maketitle

\section{Introduction}
Electro-optic (EO) effect in a material is the change in the effective refractive index upon application of the electric field. This effect has been exploited for various applications such as electro-optic modulators, sensors, etc. Perovskites with empirical formula as ABO$_3$ exhibit a large EO tensor and are a choice of materials for pure phase modulation at high frequency without optical losses at 1550 nm (communication wavelength)\cite{du1998crystal,feutmba2020strong,spirin1998measurement,uchiyama2007electro,haertling1972recent} as well as non-linear photonic applications\cite{chen1989anionic,SAHOO2016299,NUFFER20003783}. The lossless property comes from a large bandgap of perovskites.  Lithium Niobate (LNO) has been the most popular perovskite material being commercially used for intensity modulation. Recently, other perovskites such as Lead Zirconium Titanate (PZT) and Barium Titanate (BTO), which exhibit Pockels coefficient of at least an order higher than LNO, have attracted attention in optics. PZT has an ABO$_3$ structure with "Pb" corresponding to A (lattice vertices) and B corresponding to alternating "Zr" and "Ti" atoms at the body center of the tetragonal lattice with "O" atom at the face center. The non-centrosymmetric structure of PZT makes it suitable for optics and opto-MEMS with a high piezoelectric coefficient (d$_{33}$$>$ 1000 pm/V) and electro-optic (E-O) effect (r$>$ 100 pm/V). The figure of merit that governs the efficiency of an electro-optic modulator based on perovskites are V$_\pi$L and the Pockels coefficient. There have been experimental papers on PZT-based EO modulators that focus on the film crystallinity and photonic device design, but not much on the strain effects that develop in PZT due to interface lattice mismatch between PZT and buffer. PZT is integrated on Si with the help of a buffer layer that introduces either a tensile or a compressive strain on the PZT film. There have been theoretical works on Pockels coefficient calculation using DFT\cite{veithen2004first,heywang2008first,veithen2005nonlinear,cabuk2012nonlinear,shih1982theoretical,zhong2019first,kimmel2019interfacial} and enhancement due to mode softening resulting from the strained lattice wherein the EO tensor is observed to increase sharply at the phonon mode softening strain value\cite{eyraud2006interpretation,heywang2008first,fredrickson2018strain,antons2005tunability,hamze2018first,paillard2019strain}. There are reports on strain generation in PZT\cite{tanaka2000lattice}. Since the PZT film deposition on Si involves a buffer layer, a theoretical estimate would enable us to vary the deposition parameter of the buffer layers, PZT film deposition, and substrate pre-treatment to attain a strain value that gives the Pockels coefficient divergence due to phonon mode softening. There are various deposition methods for PZT such as sol-gel, MOCVD, MBE, Evaporation, PLD, and Sputter\cite{feutmba2019hybrid,lee1999drying,swartz1991sol,ramamurthi1992electrical,teowee1995electro,singh2021sputter,das2001preparation,izyumskaya2006growth,sakashita1991preparation}. We choose RF Sputter deposition due to its scalability, stoichiometry control, and fast deposition rate to corroborate our theoretical model. Parameters that could potentially affect the strain developed in the deposited PZT film are RF power, and ex-situ anneal.

Here we present the DFT simulation of strain engineering using first principles and Landau-Ginzburg-Devonshire theory and its effect on rumpling and E-O tensor. We confirm the simulation finding by using RF sputter-deposited PZT film. We calculate the lattice parameter and strain developed in the film using Bragg's law and the Williamson-Hall method from the XRD spectra of the deposited PZT film on Pt layer. The findings from the XRD data are used to fabricate Si MZI loaded with PZT and characterized for EO behavior. We confirm the strain effect with $\approx$300\% increase in EO shift on strain engineering. To the best of our knowledge, this is the first work on PZT that gives a theoretical estimation as well as experimental demonstration, using film as well as photonic device fabrication, of the effect of strain engineering of PZT film on EO tensor. The paper starts with the theoretical background for DFT simulation followed by the simulation results. It is followed by a section discussing film deposition and concludes with photonic device fabrication and discussion.

\section{Theoretical discussion}
\begin{figure}[htbp]
\centering
\includegraphics[width=\linewidth]{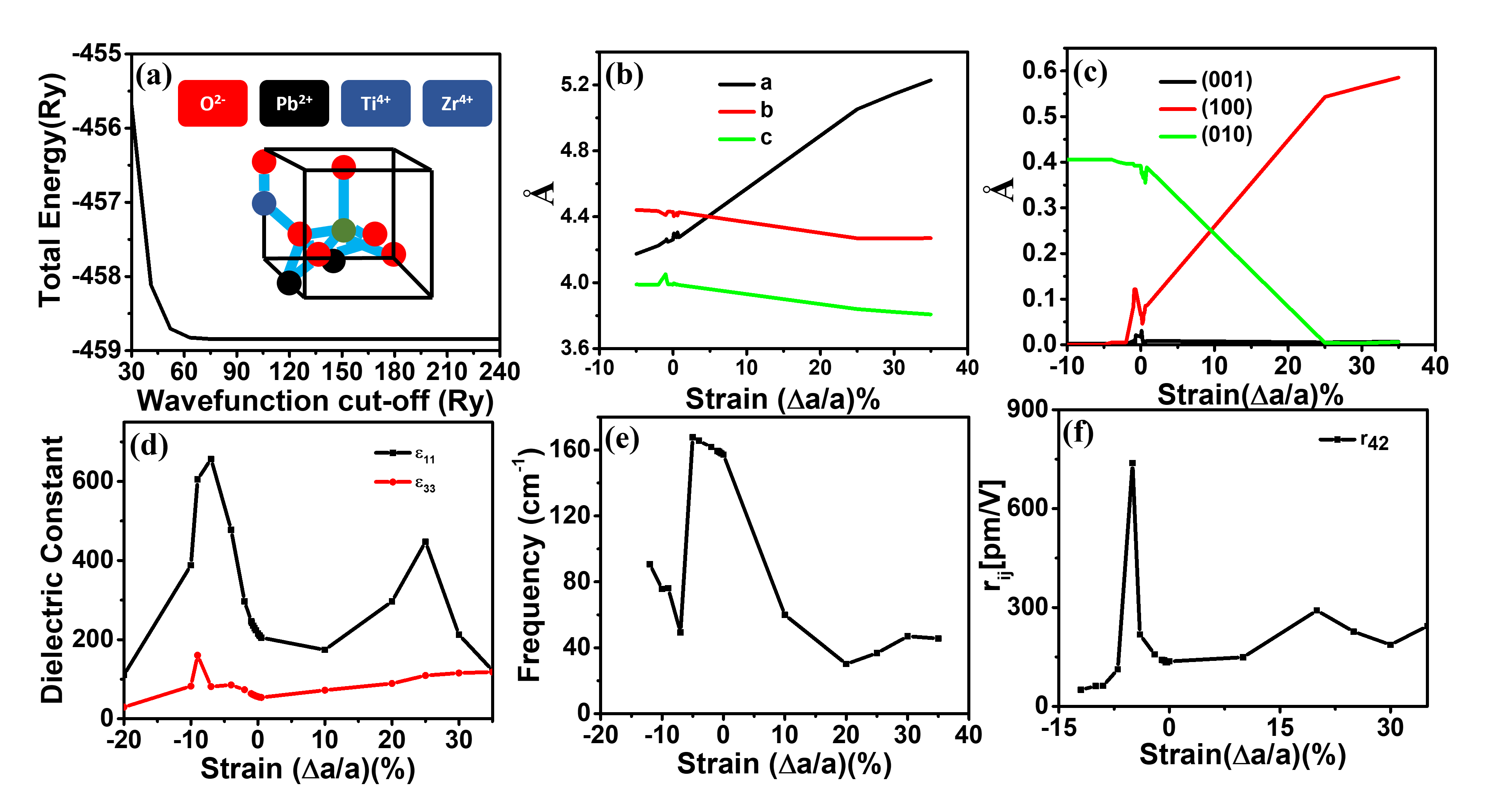}
\caption{(a)Simulation to determine the cut-off kinetic energy, Effect of variation of lattice strain on (b) lattice parameter a,b, and c, (c) rumpling in (001),(100) and (010) plane,(d) dielectric constant,(e) phonon mode softening and (f) r$_{ij}$ coefficient.}
\label{fig:Lattice_rumple}
\end{figure}
\begin{figure}[htbp]
\centering
\includegraphics[width=\linewidth]{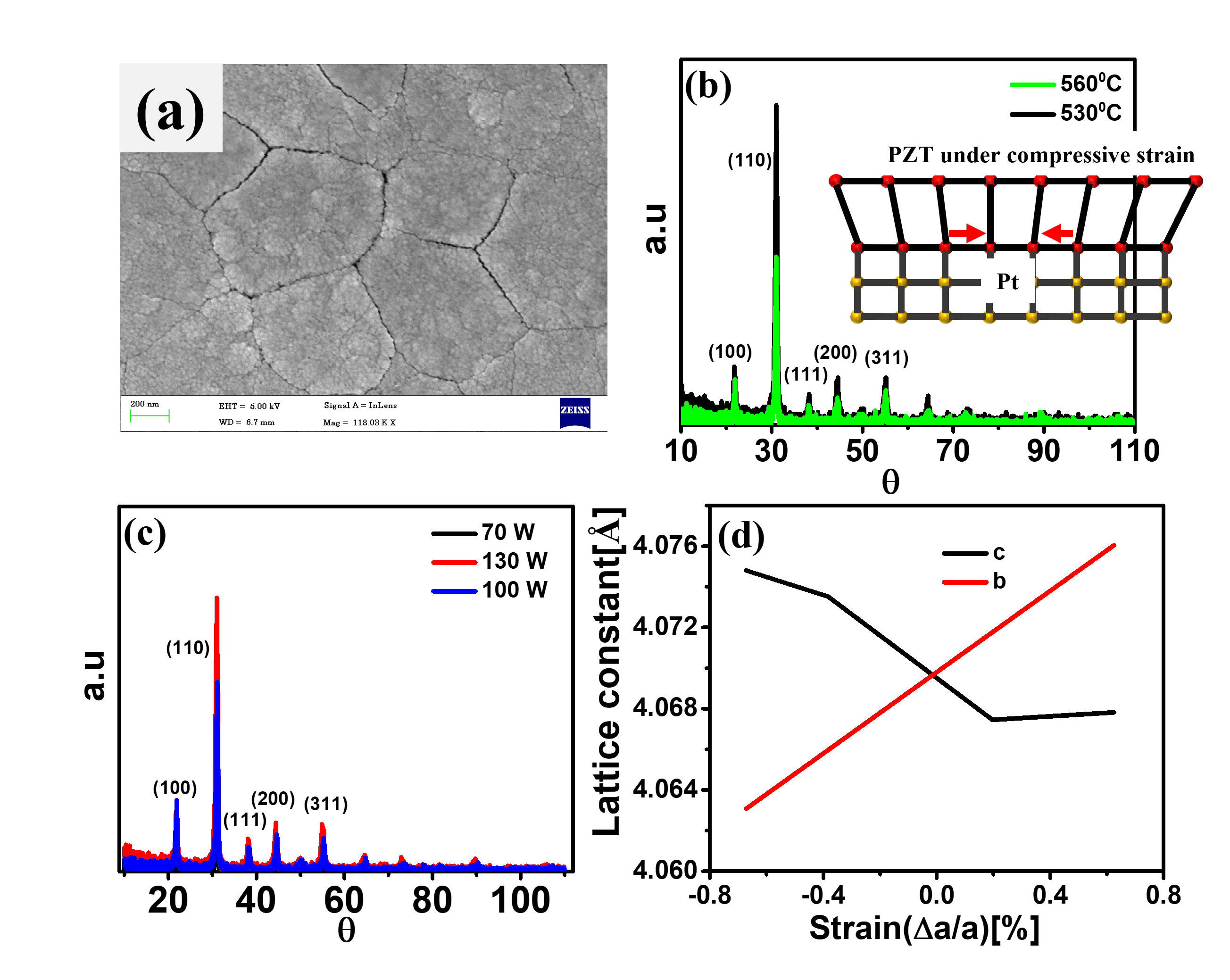}
\caption{(a)FESEM image of the fabricated PZT film at 560\degree C, 70 W RF power; XRD spectra of the deposited PZT film at varying (b) anneal temperature and (c) Deposition power;(d) variation of lattice parameter extracted from XRD spectra.}
\label{fig:Pockel_enhancement}
\end{figure}

The liner dependency of dielectric tensor $\epsilon$ on the applied electric field and linear electro-optic(E-O) tensor $r_{ijk}$ is given by eq.\ref{eq:refrac_change}. The clamped E-O tensor, with no macroscopic strain $\eta$ from the applied electric field, is a combination of electric and ionic contribution to the tensor given by eq.\ref{eq:pockel_elec} and eq.\ref{eq:pockel_ionic}.

\begin{equation}
\Delta(\epsilon^{-1})_{ij}=\sum_{k} r_{ijk}E_{k}
\label{eq:refrac_change}
\end{equation}

\begin{equation}
 r_{ijk}^{el}=-\frac{8\pi}{n^{2}_{i}n^{2}_{j}}\chi^{(2)}_{ijl}|_{l=k}
\label{eq:pockel_elec}
\end{equation}

\begin{equation}
 r_{ijk}^{el}=-\frac{4\pi}{\sqrt{\Omega}n^{2}_{i}n^{2}_{j}}\sum_{m} \frac{\alpha^{m}_{ij}p_{m,k}}{\omega^{2}_{m}}
\label{eq:pockel_ionic}
\end{equation}

$\alpha$ denotes the Raman susceptibility due to mode m with $\omega$, $\Omega$ and $p_{mk}$ denoting the transverse optic phonon frequency, unit cell volume and transverse optic mode polarities respectively .$\alpha$ is related to $\chi^{(1)}$ as in eq.\ref{eq:alpha} while $p_{mk}$ is linked to infrared intensities as in eq.\ref{eq:pmk}.

\begin{equation}
 \alpha^{m}_{ij}=\sqrt{\Omega}\sum_{\kappa,\beta}\frac{\partial\chi^{(1)}_{ij}}{\partial\tau_{\kappa,\beta}}u_{m}(\kappa\beta)
\label{eq:alpha}
\end{equation}

\begin{equation}
p_{m,k}=\sum_{\kappa,\beta}Z^{*}_{\kappa,\kappa\beta}u_{m}(\kappa\beta)
\label{eq:pmk}
\end{equation}

The net clamped E-O tensor is given by eq.\ref{eq:pockel_el_ion} 

\begin{equation}
    r^{\eta}_{ijk}=r^{el}_{ijk}+r^{ion}_{ijk}
\label{eq:pockel_el_ion}
\end{equation}

The net unclamped E-O tensor is given by eq.\ref{eq:pockel_net} wherein the contribution of the elasto-optic coefficients $p_{ij\alpha\beta}$
and the piezoelectric strain coefficients $d_{\gamma\alpha\beta}$ is taken into account. 
\begin{equation}
    r^{\sigma}_{ijk}=r^{\eta}_{ijk}+\sum^{3}_{\mu,\nu=1}p_{ij\mu\nu}d_{\kappa\mu\nu}
\label{eq:pockel_net}
\end{equation}
Landau-Ginzburg-Devonshire theory explains the relationship of large Pockels coefficient and mode softening given as eq.\ref{eq:free_energy_density}.
\begin{equation}
F(P,T)=F_{0}+\frac{1}{2}\alpha P^{2}+\frac{1}{4}\beta P^{4}
\label{eq:free_energy_density}
\end{equation}

Here, $F_{0}$,$\alpha$, and $\beta$ are energy density at zero polarization, temperature, and strain-dependent coefficients respectively with the electric field given by eq.\ref{eq:Electric Field}. 

\begin{equation}
E=\frac{\partial F}{\partial P} =\alpha P+\beta P^{3}
\label{eq:Electric Field}
\end{equation}

The linear susceptibility is the derivative of polarization with respect to the electric field given by eq.\ref{eq:linear susceptibility} and dielectric permittivity by eq.\ref{eq:permittivity} with $\chi{(2)}$ being the second derivative of polarization with respect to the electric field given in eq.\ref{eq:second_order_susceptibility}.

\begin{equation}
\chi^{(1)}=\frac{\partial P}{\partial E} =\left(\frac{\partial E}{\partial P}\right)^{-1} =\frac{1}{\alpha +3\beta P^{2}}
\label{eq:linear susceptibility}
\end{equation}

\begin{equation}
\epsilon=1+\chi^{(1)}
\label{eq:permittivity}
\end{equation}

\begin{equation}
\chi^{(2)}=\frac{\partial^{2} P}{\partial E^{2}}=-\frac{\partial^{2} E}{\partial P^{2}}\left(\frac{\partial P}{\partial E}\right)^{3}=-6\beta P (\chi^{(1)})^{3}
\label{eq:second_order_susceptibility}
\end{equation}

The dielectric tensor is related to the Pockels effect as eq.\ref{eq:r_vs_del_e} with the Pockel's tensor following the divergence shown by $\chi{^(1)}$. 

\begin{equation}
rE=\Delta\left(\frac{1}{\epsilon}\right) \approx \frac{-\Delta\epsilon}{\epsilon^{2}}=-\frac{-\chi^{(2)}E}{(1+\chi^{(1)})^{2}}=\frac{3\beta P (\chi^{(1)})^{3} E}{(1+\chi^{(1)})^{2}}
\label{eq:r_vs_del_e}
\end{equation}

As was observed in work done by Xiao-hong et.al \cite{du1998crystal} the relative permittivity is related to the stoichiometry of Zr:Ti and near to morphotropic phase boundary the permittivity $\epsilon$ increases abruptly which is orientation dependent.

\section{DFT Simulation}

We perform DFT simulation of PZT molecule using "MIT Atomic-Scale Modeling Toolkit"\cite{giannozzi2009quantum}. The calculation was done using the Monkhorst-Pack grid with GGA (Perdew-Burke-Ernzerhof) approximation using norm-conserving pseudopotential to calculate the phonon
frequencies and eigenvectors, along with the calculation of strain and EO tensor. We use the valence electron configuration of
6s$^2$5d$^{10}$6p$^2$ for Pb 3s$^2$4p$^6$5s$^2$4d$^2$ for Zr 3s$^2$3p$^6$4s$^2$3d$^2$ for Ti, and 2s$^2$2p$^4$ for O. As seen in Fig.\ref{fig:Lattice_rumple}(a), energy starts to asymptote beyond 60 Ry giving an accurate result with reduced simulation time and hence is chosen for future calculation. Fig.\ref{fig:Lattice_rumple}(b) and (c) show a cross-over between lattice parameters "a" and "b" at $\approx$5\% strain while rumpling goes to zero for strain $<$ -8\% and $>$25\% along $<$100$>$ and $<$010$>$ respectively. We observe a direct correlation between the rumpling, dielectric-tensor, and E-O tensor that diverges at the same values of strain ($\approx$400\% increase in r${_{ij}}$ at -8\% strain) as seen in Fig.\ref{fig:Lattice_rumple}(d),(e) and (f). The effect of mode softening can be seen from Fig.\ref{fig:Lattice_rumple}(e) where the frequency of phonon vibration drops at -8\% leading to an increase in the EO tensor due to $\frac{1}{\omega{^2}}$ relation leading to $\approx$400\% increase in the EO tensor from the unstrained state of PZT molecule as seen in Fig.\ref{fig:Lattice_rumple}(f). The values shown in Fig.\ref{fig:Lattice_rumple}(f) are an approximation as rumpling is not considered in all the directions giving an indication of the enhancement that can be achieved with strain engineering. To experimentally verify the effect of strain PZT was RF sputter deposited on Pt that has an $\approx$ -3\% lattice mismatch with PZT. 

\begin{figure}[htbp]
\centering
\includegraphics[width=\linewidth]{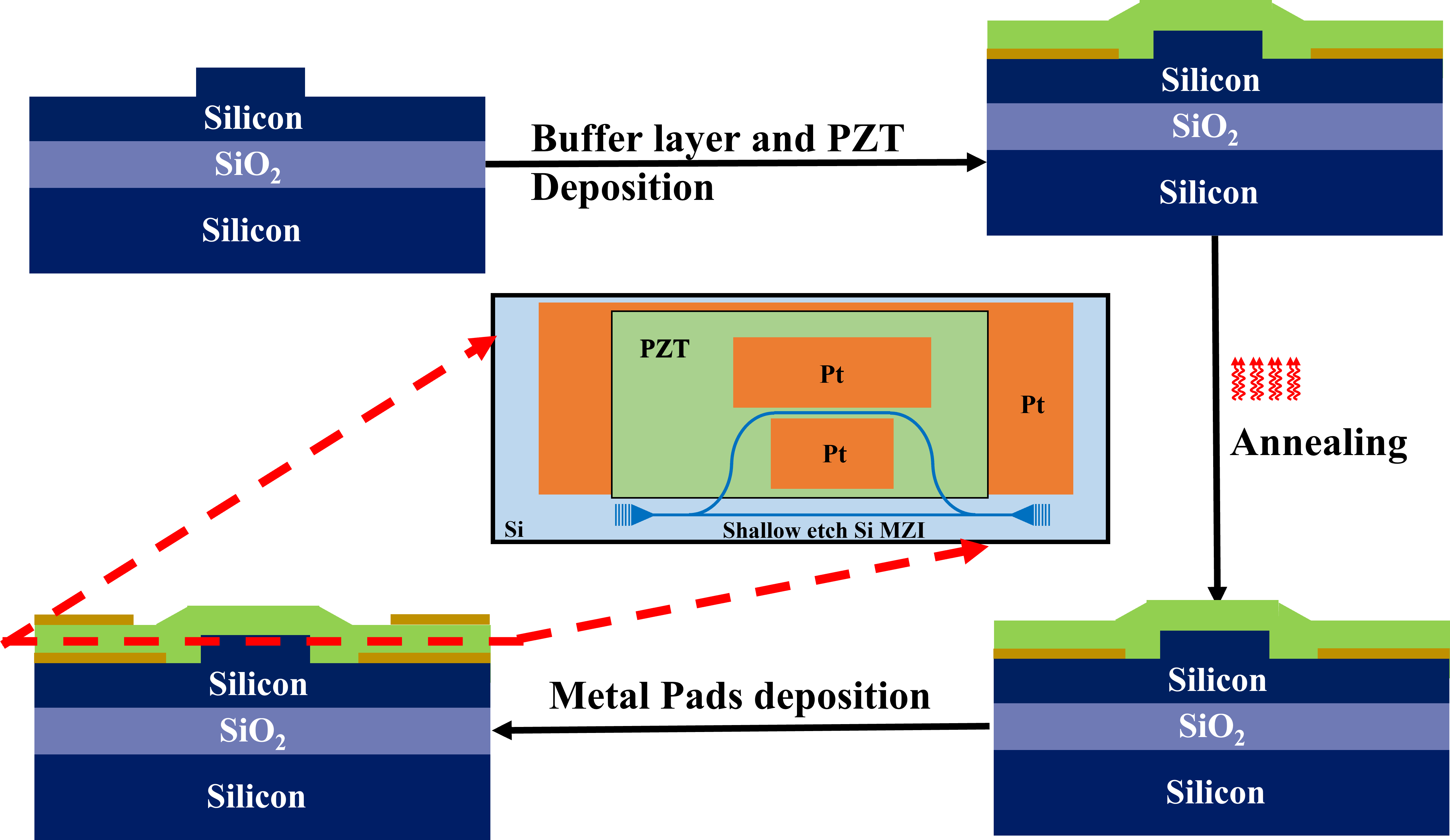}
\caption{Process flow used for Si MZI loaded with PZT EO modulator fabrication.}
\label{fig:Process_flow}
\end{figure}

\section{Experimental}
\subsection{PZT film deposition}
RF sputtering was used to deposit PZT on Pt/SiO$_2$/Si stack. Pt layer was predominantly (111) oriented. Strain variation of the deposited PZT was done by varying the deposition RF power from 70 to 130 W with an argon flow rate maintained at 30 sccm. An ex-situ anneal temperature varying from 500 to 800\degree C in an air ambiance was used to get perovskite phase. Fig.\ref{fig:Pockel_enhancement}(a) shows the FESEM image of the annealed PZT sample at 560\degree C. Fig.\ref{fig:Pockel_enhancement}(c) and (d) shows the XRD spectra for the deposited PZT at varying temperature and deposition power that was used to calculate strain developed in the film using Williamson-Hall method given by eq.\ref{eq:WHPlot}.

\begin{equation}
    \beta_T \cos(\theta)=\epsilon(4\sin(\theta))+\frac{K\lambda}{D}
    \label{eq:WHPlot}
\end{equation}

where $\beta{_T}$ is the total broadening of the peak,$\epsilon$ is strain and D is the crystallite size.
The lattice parameter is extracted from the XRD spectra using eq.\ref{eq:lattice_constant} for a tetragonal lattice.

\begin{equation}
    \frac{1}{d^2}=\frac{h^2+k^2}{a^2}+\frac{l^2}{c^2}
    \label{eq:lattice_constant}
\end{equation}

where "d" is interplanar spacing, "a", "c" are the lattice constants, and (hkl) are the miller indices. "d" is calculated using Bragg's law given by

\begin{equation}
    d=\frac{n\lambda}{2\sin(\theta)}
\end{equation}

where $\theta$ is the peak positions in XRD spectra.

\begin{table}[t]
\centering
\caption{\bf Strain calculation for deposited PZT film}
\begin{tabular}{ccc}
\hline
RF Power & Anneal Temperature & Strain(\%) \\
\hline
70 W & 530\degree C & -0.21 \\
70 W & 560\degree C & -0.14\\
70 W & 600\degree C & -0.04\\
100 W & 560\degree C & -0.13\\
130 W & 560\degree C & -0.13\\
\hline
\end{tabular}
  \label{tab:strain_calculation}
\end{table}

\begin{figure}[htbp]
\centering
\includegraphics[width=\linewidth]{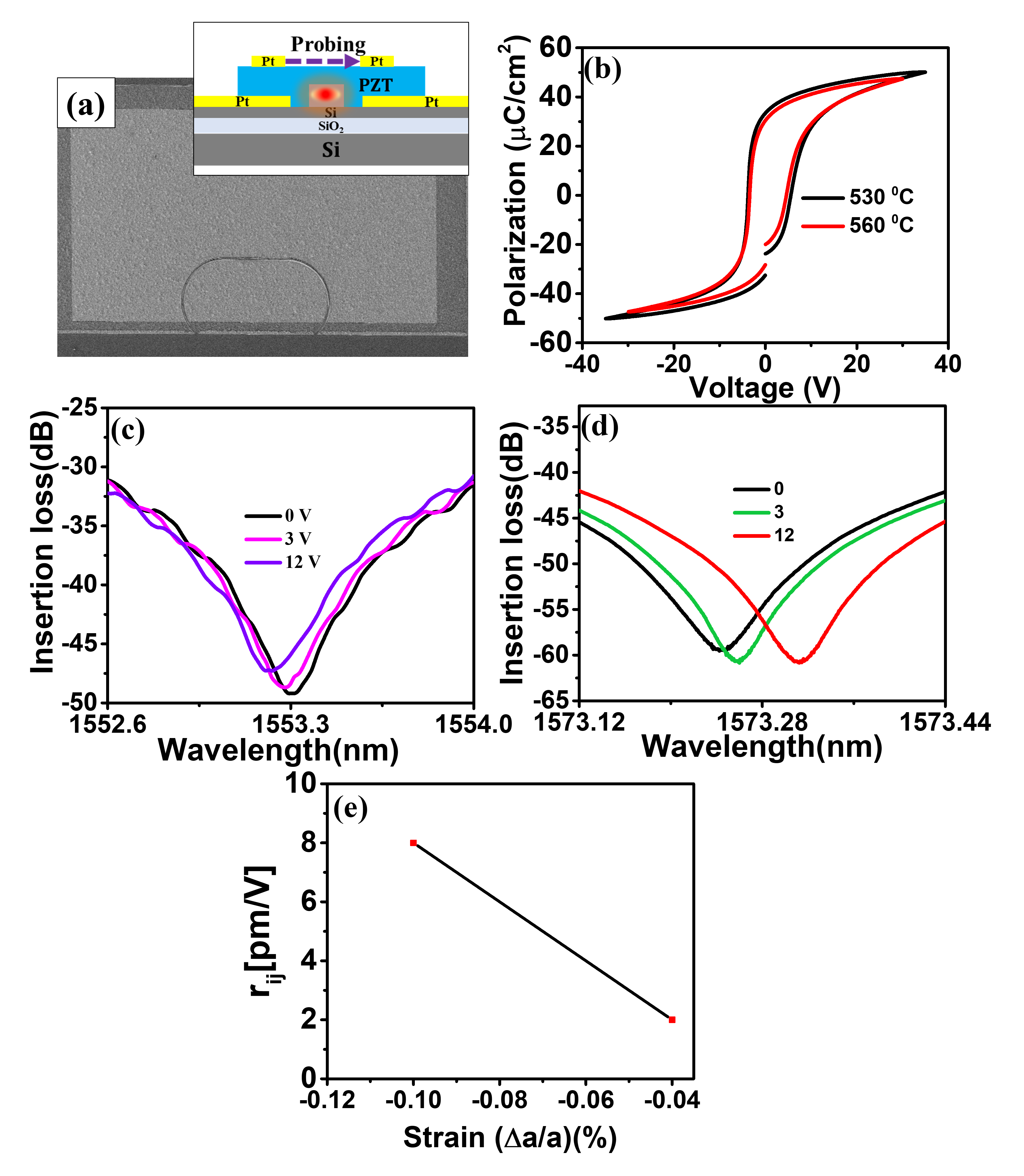}
\caption{(a) FESEM and cross-section schematic of the fabricated device;(b) P-V loop of the deposited PZT film annealed at 530 and 560\degree C; EO shift on voltage application for (c) 600\degree C with -0.04\% strain and (d) 530\degree C with -0.21\% strain;(e) variation of measured Pockels coefficient with strain.}
\label{fig:Proposed_device}
\end{figure}

Table.\ref{tab:strain_calculation} shows that the annealing temperature controls the strain predominantly as compared to the RF deposition power used with the film most strained at 530\degree C. The trend in the simulated lattice parameter variation as shown in Fig.\ref{fig:Lattice_rumple}(b) matches with the experimental observation shown in Fig.\ref{fig:Pockel_enhancement}(d) with the shift in the "a" parameter being attributed to lattice relaxation allowed in simulation while the cell structure is assumed fixed for experimental calculation. 

\subsection{Photonic device fabrication and characterization}
The experimental confirmation of the Pockels coefficient enhancement was done by making Si Mach-Zhender Interferometer(MZI) loaded with PZT along one arm with the probing done using co-planar electrodes. Fig.\ref{fig:Process flow} shows the process flow of the fabricated MZI device. A 90 nm shallow etch Si MZI is fabricated on an SOI substrate. It is followed by DC sputter deposition of Ti/Pt layer with a thickness of 20/70 nm. It is annealed in air ambiance at 650\degree C to obtain a predominantly (111) oriented Pt layer. PZT is RF sputter deposited at 70 W RF power with a 30 sccm argon flow rate and source-target distance of 6 cm with a deposition time of 2Hr. The deposited PZT film is air annealed at varying temperatures of 530\degree C, 560\degree C, and 600\degree C. Fig.\ref{fig:Proposed_device}(a) shows the FESEM image and the schematic of the fabricated device with Si acting as the waveguiding medium and PZT as an optically active material. Fig.\ref{fig:Proposed_device}(b) shows the polarization-Voltage(P-V) curve for the deposited film at 530\degree C and 560\degree C with 530\degree C annealed film showing higher polarization than film annealed at 560\degree C. Fig.\ref{fig:Proposed_device}(c) and (d) correspond to PZT annealed at 600 and 530\degree C respectively with a DC optical spectrum shift of 2 pm/V and 8 pm/V respectively. This corresponds to an enhancement of electro-optic shift of $\approx$300\% corresponding to strain change from -0.04\% to -0.21\% from Table.\ref{tab:strain_calculation} as seen in Fig.\ref{fig:Proposed_device}(e).

\section{Conclusion}
We report $\approx$400\% enhancement in PZT Pockels coefficient on DFT simulation of lattice strain due to phonon mode softening. The simulation showed a relation between the rumpling and the Pockels coefficient divergence at a strain of -8\% and 25\%. The simulation was verified experimentally by RF sputter deposited PZT film on Pt/SiO$_2$/Si layer. The strain developed in PZT varied from -0.04\% for film annealed at 530\degree C to -0.21\% for 600\degree C annealing temperature. The strain was insensitive to RF power with a value of -0.13\% for power varying between 70-130 W. Pockels coefficient enhancement was experimentally confirmed by Si Mach–Zehnder interferometer (MZI) loaded with PZT and probed with the co-planar electrode. An enhancement of $\approx$300\% in Pockels coefficient was observed from 2-8 pm/V with strain increasing from -0.04\% to -0.21\%. To the best of our knowledge, this is the first time studying and demonstration of strain engineering on Pockels coefficient of PZT using DFT simulation, film deposition, and photonic device fabrication.

\section*{Acknowledgments}
SKS thanks Professor Ramakrishna Rao chair fellowship.

\section*{Disclosures}
The authors declare no conflicts of interest.

\section*{Data availability}
Data underlying the results presented in this paper are not publicly available at this time but may be obtained from the authors upon reasonable request.

%Bibliography
\bibliographystyle{unsrt}  
\bibliography{biblio_file}

\begin{thebibliography}{10}

\bibitem{du1998crystal}
Xiao-hong Du, Jiehui Zheng, Uma Belegundu, and Kenji Uchino.
\newblock Crystal orientation dependence of piezoelectric properties of lead
  zirconate titanate near the morphotropic phase boundary.
\newblock {\em Applied physics letters}, 72(19):2421--2423, 1998.

\bibitem{feutmba2020strong}
Gilles~Freddy Feutmba, Tessa Van~de Veire, Irfan Ansari, John~P George, Dries
  Van~Thourhout, and Jeroen Beeckman.
\newblock A strong pockels pzt/si modulator for efficient electro-optic tuning.
\newblock In {\em Integrated Photonics Research, Silicon and Nanophotonics},
  pages ITu1A--6. Optica Publishing Group, 2020.

\bibitem{spirin1998measurement}
Vasilii~V Spirin, Changho Lee, and Kwangsoo No.
\newblock Measurement of the pockels coefficient of lead zirconate titanate
  thin films by a two-beam polarization interferometer with a reflection
  configuration.
\newblock {\em JOSA B}, 15(7):1940--1946, 1998.

\bibitem{uchiyama2007electro}
Kiyoshi Uchiyama, Atsushi Kasamatsu, Yohei Otani, and Tadashi Shiosaki.
\newblock Electro-optic properties of lanthanum-modified lead zirconate
  titanate thin films epitaxially grown by the advanced sol--gel method.
\newblock {\em Japanese journal of applied physics}, 46(3L):L244, 2007.

\bibitem{haertling1972recent}
GH~Haertling and CE~Land.
\newblock Recent improvements in the optical and electrooptic properties of
  plzt ceramics.
\newblock {\em Ferroelectrics}, 3(1):269--280, 1972.

\bibitem{chen1989anionic}
Chuangtian Chen, Yicheng Wu, and Rukang Li.
\newblock The anionic group theory of the non-linear optical effect and its
  applications in the development of new high-quality nlo crystals in the
  borate series.
\newblock {\em International Reviews in Physical Chemistry}, 8(1):65--91, 1989.

\bibitem{SAHOO2016299}
M.P.K. Sahoo, Yajun Zhang, and Jie Wang.
\newblock Enhancement of ferroelectric polarization in layered bazro3/batio3
  superlattices.
\newblock {\em Physics Letters A}, 380(1):299--303, 2016.

\bibitem{NUFFER20003783}
J~Nuffer, D.C Lupascu, and J~Rödel.
\newblock Damage evolution in ferroelectric pzt induced by bipolar electric
  cycling.
\newblock {\em Acta Materialia}, 48(14):3783--3794, 2000.

\bibitem{veithen2004first}
Marek Veithen, Xavier Gonze, and Philippe Ghosez.
\newblock First-principles study of the electro-optic effect in ferroelectric
  oxides.
\newblock {\em Physical review letters}, 93(18):187401, 2004.

\bibitem{heywang2008first}
Walter Heywang, Karl Lubitz, Wolfram Wersing, and RE~Cohen.
\newblock First-principles theories of piezoelectric materials.
\newblock {\em Piezoelectricity: Evolution and Future of a Technology}, pages
  471--492, 2008.

\bibitem{veithen2005nonlinear}
M~Veithen, Xavier Gonze, and Ph~Ghosez.
\newblock Nonlinear optical susceptibilities, raman efficiencies, and
  electro-optic tensors from first-principles density functional perturbation
  theory.
\newblock {\em Physical Review B}, 71(12):125107, 2005.

\bibitem{cabuk2012nonlinear}
Suleyman Cabuk.
\newblock The nonlinear optical susceptibility and electro-optic tensor of
  ferroelectrics: first-principle study.
\newblock {\em Central European Journal of Physics}, 10:239--252, 2012.

\bibitem{shih1982theoretical}
Chun-Ching Shih and Amnon Yariv.
\newblock A theoretical model of the linear electro-optic effect.
\newblock {\em Journal of Physics C: Solid State Physics}, 15(4):825, 1982.

\bibitem{zhong2019first}
Mi~Zhong, Wei Zeng, Fu-Sheng Liu, Bin Tang, and Qi-Jun Liu.
\newblock First-principles study of the atomic structures, electronic
  properties, and surface stability of batio3 (001) and (011) surfaces.
\newblock {\em Surface and Interface Analysis}, 51(10):1021--1032, 2019.

\bibitem{kimmel2019interfacial}
Anna~V Kimmel.
\newblock Interfacial phenomena in nanocapacitors with multifunctional oxides.
\newblock {\em Physical Chemistry Chemical Physics}, 21(44):24643--24649, 2019.

\bibitem{eyraud2006interpretation}
Lucien Eyraud, Benoit Guiffard, Laurent Lebrun, and Daniel Guyomar.
\newblock Interpretation of the softening effect in pzt ceramics near the
  morphotropic phase boundary.
\newblock {\em Ferroelectrics}, 330(1):51--60, 2006.

\bibitem{fredrickson2018strain}
Kurt~D Fredrickson, Viola~Valentina Vogler-Neuling, Kristy~J Kormondy, Daniele
  Caimi, Felix Eltes, Marilyne Sousa, Jean Fompeyrine, Stefan Abel, and
  Alexander~A Demkov.
\newblock Strain enhancement of the electro-optical response in bati o 3 films
  integrated on si (001).
\newblock {\em Physical Review B}, 98(7):075136, 2018.

\bibitem{antons2005tunability}
Armin Antons, JB~Neaton, Karin~M Rabe, and David Vanderbilt.
\newblock Tunability of the dielectric response of epitaxially strained srti o
  3 from first principles.
\newblock {\em Physical Review B}, 71(2):024102, 2005.

\bibitem{hamze2018first}
Ali~K Hamze and Alexander~A Demkov.
\newblock First-principles study of the linear electro-optical response in
  strained srti o 3.
\newblock {\em Physical Review Materials}, 2(11):115202, 2018.

\bibitem{paillard2019strain}
Charles Paillard, Sergei Prokhorenko, and Laurent Bellaiche.
\newblock Strain engineering of electro-optic constants in ferroelectric
  materials.
\newblock {\em npj Computational Materials}, 5(1):6, 2019.

\bibitem{tanaka2000lattice}
Keisuke Tanaka, Yoshiaki Akiniwa, Yoshihisa Sakaida, and Hirohisa Kimachi.
\newblock Lattice strain and domain switching induced in tetragonal pzt by
  poling and mechanical loading.
\newblock {\em JSME International Journal Series A Solid Mechanics and Material
  Engineering}, 43(4):351--357, 2000.

\bibitem{feutmba2019hybrid}
Gilles~F Feutmba, John~P George, Koen Alexander, Dries Van~Thourhout, and
  Jeroen Beeckman.
\newblock Hybrid pzt/si tm/te electro-optic phase modulators.
\newblock In {\em Integrated Optics: Devices, Materials, and Technologies
  XXIII}, volume 10921, pages 85--91. SPIE, 2019.

\bibitem{lee1999drying}
Changho Lee, Vasili Spirin, Hanwook Song, and Kwangsoo No.
\newblock Drying temperature effects on microstructure, electrical properties
  and electro-optic coefficients of sol-gel derived pzt thin films.
\newblock {\em Thin Solid Films}, 340(1-2):242--249, 1999.

\bibitem{swartz1991sol}
SL~Swartz, SD~Ramamurthi, JR~Busch, and VE~Wood.
\newblock Sol-gel pzt films for optical waveguides.
\newblock {\em MRS Online Proceedings Library (OPL)}, 243:533, 1991.

\bibitem{ramamurthi1992electrical}
SD~Ramamurthi, SL~Swartz, KR~Marken, JR~Busch, and VE~Wood Battelle.
\newblock Electrical and optical properties of sol-gel processed pb (zr, ti) o3
  films.
\newblock {\em MRS Online Proceedings Library (OPL)}, 271, 1992.

\bibitem{teowee1995electro}
G~Teowee, JT~Simpson, Tianji Zhao, M~Mansuripur, JM~Boulton, and DR~Uhlmann.
\newblock Electro-optic properties of sol-gel derived pzt and plzt thin films.
\newblock {\em Microelectronic engineering}, 29(1-4):327--330, 1995.

\bibitem{singh2021sputter}
Suraj Singh and Shankar~Kumar Selvaraja.
\newblock Sputter-deposited pzt-on-silicon electro-optic modulator.
\newblock In {\em 2021 IEEE Photonics Conference (IPC)}, pages 1--2. IEEE,
  2021.

\bibitem{das2001preparation}
RN~Das, A~Pathak, SK~Saha, S~Sannigrahi, and P~Pramanik.
\newblock Preparation, characterization and property of fine pzt powders from
  the poly vinyl alcohol evaporation route.
\newblock {\em Materials research bulletin}, 36(9):1539--1549, 2001.

\bibitem{izyumskaya2006growth}
Natalia Izyumskaya, Vitaliy Avrutin, Xing Gu, Umit Ozgur, Bo~Xiao, Tae~Dong
  Kang, Hosun Lee, and Hadis Morkoc.
\newblock Growth of high-quality pb (zrxti1-x) o3 films by peroxide mbe and
  their optical and structural characteristics.
\newblock {\em MRS Online Proceedings Library (OPL)}, 966, 2006.

\bibitem{sakashita1991preparation}
Yukio Sakashita, Toshiyuki Ono, Hideo Segawa, Kouji Tominaga, and Masaru Okada.
\newblock Preparation and electrical properties of mocvd-deposited pzt thin
  films.
\newblock {\em Journal of applied physics}, 69(12):8352--8357, 1991.

\bibitem{giannozzi2009quantum}
Paolo Giannozzi, Stefano Baroni, Nicola Bonini, Matteo Calandra, Roberto Car,
  Carlo Cavazzoni, Davide Ceresoli, Guido~L Chiarotti, Matteo Cococcioni,
  Ismaila Dabo, et~al.
\newblock Quantum espresso: a modular and open-source software project for
  quantum simulations of materials.
\newblock {\em Journal of physics: Condensed matter}, 21(39):395502, 2009.

\end{thebibliography}

\end{document}